\documentclass[12pt]{article} 
\newcommand{\be}{\begin{equation}}
\newcommand{\ee}{\end{equation}}
\newcommand{\ba}{\begin{array}}
\newcommand{\ea}{\end{array}}

\newcommand{\dvg}{{\rm div\;}}
\newcommand{\grad}{{\rm \; grad\;}}
\newcommand{\curl}{{\rm \; curl\;}}

\newcommand{\RR}{\mathbb R}

\newcommand{\bea}{\begin{eqnarray*}}
\newcommand{\eea}{\end{eqnarray*}}
\newcommand{\bean}{\begin{eqnarray}}
\newcommand{\eean}{\end{eqnarray}}
\newcommand{\proof}{\vspace{1ex}\noindent{\em Proof}. \ }
\def\ds{\displaystyle}
\def\nm{\noalign{\medskip}}

\newcommand\eps{\varepsilon}

\newtheorem{lemma}{Lemma}[section]

\newtheorem{definition}{Definition}[section]
\newtheorem{theorem}{Theorem}[section]

\newtheorem{proposition}{Proposition}[section]
\usepackage{amsmath, amssymb}
\usepackage{latexsym}
\newcommand{\R}{\mathbb{R}}

\def\Box{\leavevmode\vbox{\hrule
     \hbox{\vrule\kern5pt\vbox{\kern5pt}%
           \vrule}\hrule}}
\renewcommand{\square}{\hfill$\Box$}
\begin{document}
\title{Recovery of small electromagnetic inhomogeneities from boundary
measurements in time-dependent Maxwell's equations}

%Identification of inhomogeneities with small diameter by boundary
%measurements for the time-dependent Maxwell's
 % equations}
\author{Christian Daveau \thanks{
D\'epartement de Math\'ematiques, Site Saint-Martin II, BP 222, \&
Universit\'e de Cergy-Pontoise, 95302 Cergy-Pontoise Cedex, France
(Email: christian.daveau@math.u-cergy.fr).}   Abdessatar Khelifi
\thanks{ D\'epartement de Math\'ematiques \& Informatique
Facult\'e des Sciences, 7021 Zarzouna - Bizerte, Tunisia
(Email:abdessatar.khelifi@fsb.rnu.tn).}}
%\date{}
\maketitle \abstract{ We consider for the time-dependent Maxwell's
equations the inverse problem of identifying locations
 and certain properties of small electromagnetic inhomogeneities
in a homogeneous background medium from dynamic measurements of
the tangential component of the magnetic field on the boundary (
or a part of the boundary) of a domain.}\\



\section{Introduction}
Let $\Omega$ be a bounded $C^2$-domain in $\mathbb{R}^d$, $d=
2,3$. Assume  that $\Omega$ contains a finite number of
inhomogeneities, each of the form $z_j + \alpha B_j$, where $B_j
\subset \R^d$ is a bounded, smooth domain containing the origin.
The total collection of inhomogeneities is
 $ \ds {\cal B}_\alpha
 = \ds \cup_{j=1}^{m} (z_j  + \alpha B_j)$.
  The points $z_j \in \Omega, j=1, \ldots, m,$ which determine the
  location of the inhomogeneities, are assumed to satisfy the
  following inequalities:
\be \label{f1}
 | z_j  - z_{j^\prime} | \geq c_0 > 0, \forall \; j \neq {j^\prime}  \quad
\mbox{ and } \mbox{ dist} (z_j, \partial \Omega) \geq c_0
> 0, \forall \; j.
\ee Assume that $\alpha
>0$, the common order of magnitude of the diameters of the
inhomogeneities, is sufficiently small, that these
 inhomogeneities are disjoint,  and that
their distance to $\R^d \setminus \overline{\Omega}$ is larger
than $c_0/2$.  Let $\mu_0$ and $\eps_0$ denote the permeability
and the permittivity of the background medium, and assume that
$\mu_0>0$ and $\eps_0>0$ are positive constants.  Let $\mu_j>0$
and $\eps_j>0$ denote the permeability and the permittivity of the
j-th inhomogeneity, $z_j+\alpha B_j$, these are also assumed to be
positive constants. Introduce the piecewise-constant electric
permittivity
\begin{equation}
\eps_\alpha(x)=\left \{ \begin{array}{*{2}{l}}
 \eps_0,\;\;& x \in \Omega \setminus \bar {\cal B}_\alpha,  \\
 \eps_j,\;\;& x \in z_j+\alpha B_j, \;j=1 \ldots m.
\end{array}
\right . \label{murhodef}
\end{equation}
If we allow the degenerate case $\alpha =0$, then the function
$\eps_0(x)$ equals the constant $\eps_0$. Assume that, the
magnetic permeability is given by
\begin{equation}\label{assump1}
\mu_\alpha(x)=\mu_0,\quad  \mbox{ for all }x\in
\Omega.\end{equation} Let $\nu =\nu (x)$ denote the outward unit
normal vector to $\Omega$ at a point on $\partial \Omega$, and
$\partial_t=
\frac{\partial}{\partial t}$. \\

In this paper, we will denote by bold letters the functional
spaces for the vector fields. Thus $H^{s}(\Omega)$ denotes the
usual Sobolev space on
 $\Omega$ and ${\bf H}^s(\Omega)$ denotes $(H^s(\Omega))^d$ and ${\bf L}^2(\Omega)$
  denotes $(L^2(\Omega))^d$. As usual for Maxwell
 equations, we need spaces of fields with square integrable curls:
\[
{\bf H}(\curl; \Omega) = \{ u \in {\bf L}^2(\Omega), \curl u \in
{\bf L}^2(\Omega) \},\] and with square integrable divergences
\[
{\bf H}(\dvg; \Omega) = \{ u \in {\bf L}^2(\Omega), \dvg u \in
 L^2(\Omega) \}.\] We will also need the following functional
spaces:
\[
Y(\Omega) = \{ u \in {\bf L}^2(\Omega), \dvg u = 0 {\rm \; in\;}
\Omega
 \},\quad X(\Omega) = {\bf H}^1(\Omega)\cap Y(\Omega),
\] and $TL^2(\partial \Omega)$ the space of vector fields on
$\partial \Omega$
  that lie in ${\bf L}^2(\partial\Omega)$. Finally, the "minimal" choice for
   the magnetic variational space would be
   \[
 X_N(\Omega)=\{v\in {\bf H}(\curl; \Omega)\cap {\bf H}(\dvg; \Omega) ;\quad
 v\times \nu=0\quad \mbox{on } \partial\Omega\}. \]

We use $\langle\cdot,\cdot\rangle$ for the duality bracket and
$(\cdot,\cdot)$ for the ${\bf L}^2$ product.\\
%%%%%%%%%%%%%%%%%%%%%%%%%%%%%%%%%%%%%%%%%%%%%%%%%%%%%%%%%%%%%%%%%%%%%%%%%%%%%%%%%%

In the non flawed region $\Omega$ an electric field/magnetic field
pair $(E,H)$ satisfies: \be \label{eq-Max0} \left\{
\begin{array}{l}
\ds \curl E = -\mu_0\partial_t H \quad {\rm in}\;
\Omega \times (0, T),\\
\ds \curl H = \eps_0\partial_t E  \quad {\rm in}\; \Omega \times
(0, T),
\end{array}
\right. \ee

Let $TH_{\dvg}^{-1/2}(\partial\Omega)$ denote the space of
tangential vector fields on $\partial\Omega$ that lie in
$H^{-1/2}(\partial\Omega)$. The most common boundary data is \be
(E \times \nu) |_{\partial \Omega \times (0, T)}\quad \mbox{ given
in  }  TH_{\dvg}^{-1/2}(\partial\Omega).\ee By dividing the second
equation in (\ref{eq-Max0}) by $\eps_0$ and taking the curl, we
obtain in terms of the magnetic field: \be\label{wom} \ds
 \curl (\frac{1}{\eps_0}\curl H) +\mu_0\partial_t^2H= 0 \quad {\rm in}\;
\Omega \times (0, T), \ee and the boundary data is supposed to be
given by \be ( H \times \nu) |_{\partial \Omega \times (0, T)}=f
\quad \mbox{ given in } TH_{\dvg}^{-1/2}(\partial\Omega).\ee
Moreover, we set \be H |_{t=0} = \varphi, \partial_t H|_{t=0} =
\psi \quad {\rm in}\; \Omega.\ee
 Here $T>0$ is a final observation time and
 $\varphi, \psi \in {\cal
C}^{\infty}(\overline{\Omega})$
 and $f \in {\cal C}^{\infty}(0, T; {\cal C}^{\infty}(\partial
 \Omega))$ are subject to the compatibility conditions
\[ \partial_{t}^{2l} f |_{t = 0} =  (\Delta^l \varphi) \times \nu
|_{\partial \Omega} {\rm \; and \;} \partial_t^{2l + 1}  f |_{t
=0} =  (\Delta^l \psi ) \times \nu |_{\partial \Omega}, \quad l=1,
2, \ldots \]

Let $H_\alpha\in \RR^d$ be the magnetic field corresponding to the
case of the presence of a finite number of small electromagnetic
inhomogeneities. This field (under the assumption (\ref{assump1}))
satisfies \be \label{alphwam} \left\{
\begin{array}{l}
\curl (\frac{1}{\eps_\alpha} \curl  H_\alpha) + \mu_0\partial_t^2
H_\alpha= 0 \quad
{\rm in}\; \Omega \times (0, T),\\
\nm \dvg (\mu_0 H_\alpha) = 0 \quad
{\rm in}\; \Omega \times (0, T),\\
\nm H_\alpha |_{t=0} = \varphi,
\partial_t H_\alpha |_{t=0} = \psi \quad {\rm in}\; \Omega,\\ \nm
H_\alpha \times \nu |_{\partial \Omega \times (0, T)} = f,
\end{array}
\right. \ee It is well known that (\ref{wom}) has a unique
solution $ H \in {\cal C}^\infty ([0, T] \times
\overline{\Omega})$. It is also known (see for example \cite{NI})
that since $\Omega$ is smooth (${\cal C}^2-$ regularity would be
sufficient) the non homogeneous Maxwell's equations
(\ref{alphwam}) have a unique weak solution $H_\alpha \in {\cal
C}^0(0, T; X(\Omega)) \cap {\cal C}^1(0, T; {\bf L}^2(\Omega))$.
Indeed, $\curl H_\alpha$ belongs to $ {\cal C}^0(0, T; X(\Omega))
\cap {\cal C}^1(0, T; {\bf L}^2(\Omega))$.\\

Having found $H_\alpha$, we then obtain the field $E_\alpha$
through the formula:
\[
\partial_t E_\alpha=\ds\frac{1}{\eps_\alpha} \curl H_\alpha.
\]
Our main goal in this paper is to determine, most effectively,
properties of the inhomogeneities $z_j + \alpha B_j$, from over
determined boundary information about specific solutions to
(\ref{alphwam}). In particular, we study media that consist of a
homogeneous(constant coefficient) electromagnetic material with a
finite number of small inhomogeneities, and as our main result we
derive asymptotic formulas for the perturbations in the
(tangential) boundary magnetic fields caused by the presence of
these inhomogeneities. Our formulas may be used to determine
properties (location, relative size) of the small inhomogeneities
in case a single, or a few (tangential) boundary electric fields
and their corresponding(tangential)boundary magnetic fields are
known. For stationary Maxwell's equations it has been known that
the Dirichlet to Neumann map uniquely determines (smooth)
isotropic electromagnetic parameters, see \cite{M2}, \cite{RK},
\cite{SIC}. We will provide in this paper a rigorous derivation of
the inverse Fourier transform of a linear combination of
derivatives of point masses, located at the positions $z_j$ of the
inhomogeneities, as the leading order term of an appropriate
averaging of (partial) dynamic boundary measurements of the
tangential components of magnetic fields on part of the boundary.
We refer the reader to \cite{SU},\cite{VV}, \cite{CMV}, and
\cite{FV} for discussions on closely related (stationary)
identification problems.\\

Our approach, aimed at determining specific internal features of
an object based on electromagnetic boundary measurements, differs
from \cite{A}, \cite{A-al}, \cite{AMV}, \cite{AVV}, \cite{Volkov}
and it can be regarded as constructive method, but until tested.

%%%%%%%%%%%%%%%%%%%%%%%%%%%%%%%%%%%%%%%%%%%%%%%%%%%%%%%%%%%%%%%%%%%%%%%%%%%%%%%%%%%%%%
\section{Asymptotic behavior}
We start the derivation of the asymptotic formula for $\ds \curl
H_\alpha\times \nu$ with the following estimate.

\begin{lemma}\label{est-lem1}
The following estimate as $\alpha \rightarrow 0$ holds: \be
\label{estenergy0} || \partial_t (H_\alpha - H) ||_{L^{\infty}(0,
T; {\bf L}^2(\Omega))} + || H_\alpha - H ||_{L^{\infty}(0, T;
X_N(\Omega))}  \leq C \alpha^d, \ee
 where the constant $C$ is independent of
$\alpha$ and the set of points $\{  z_j\}_{j=1}^m$ provided that
assumption (\ref{f1}) holds.
\end{lemma}
\proof From (\ref{wom})-(\ref{alphwam}), it is obvious that
$H_\alpha - H \in X_N(\Omega)$, then due to the Green formula we
have for any ${\bf v} \in X_N(\Omega)$: \be \label{eq2} \ds
\int_{\Omega} \mu_0
\partial_t^2 (H_\alpha -H) \cdot{\bf v} + \ds\int_{\Omega}
\frac{1}{\eps_\alpha} \curl (H_\alpha -H) \cdot \curl {\bf v} =
\sum_{j=1}^m (\frac{1}{\eps_0} - \frac{1}{\eps_j}) \int_{z_j +
\alpha B_j} \curl H \cdot \curl {\bf v}. \ee Let ${\bf v}_\alpha$
be defined by \be\label{rel-lem1-1} \left\{
\begin{array}{l}
{\bf v}_\alpha \in X_N(\Omega),\\
\nm \curl \frac{1}{\eps_\alpha} \curl {\bf v}_\alpha = \partial_t
(H_\alpha - H) \quad {\rm in\;} \Omega.
\end{array}
\right. \ee Then,
\[
\ds  \int_{\Omega} \frac{1}{\eps_\alpha} \curl (H_\alpha -H) \cdot
\curl {\bf v}_\alpha  = - \int_{\Omega} \partial_t (H_\alpha - H)
\cdot(H_\alpha -H) = - \frac{1}{2} \partial_t \int_{\Omega}
|H_\alpha - H|^2
\]
and by Green formula, relation (\ref{rel-lem1-1}) gives:
\[
\begin{array}{lll}
\ds \int_{\Omega} \partial_t^2 (H_\alpha -H)\cdot {\bf v}_\alpha
&=& \ds \int_{\Omega}
\curl \frac{1}{\eps_\alpha} \curl \partial_t {\bf v}_\alpha \cdot {\bf v}_\alpha\\
\nm &=& \ds - \int_{\Omega}
\frac{1}{\eps_\alpha} \curl \partial_t {\bf v}_\alpha \cdot \curl {\bf v}_\alpha\\
\nm &=& \ds - \frac{1}{2} \partial_t \int_{\Omega}
\frac{1}{\eps_\alpha} |\curl {\bf v}_\alpha|^2.\end{array}
\]
Thus, it follows from (\ref{eq2}) that
\[
\ds \mu_0\partial_t \int_{\Omega} \frac{1}{\eps_\alpha} |\curl
{\bf v}_\alpha|^2 +
\partial_t \int_{\Omega} |H_\alpha - H|^2 =
 -2 \sum_{j=1}^m (\frac{1}{\eps_0} - \frac{1}{\eps_j}) \int_{z_j
 + \alpha B_j} \curl H \cdot \curl {\bf v}_\alpha.
\]
Next,
\[
\ds | \sum_{j=1}^m (\frac{1}{\eps_0} - \frac{1}{\eps_j})
 \int_{z_j + \alpha B_j} \curl H \cdot \curl {\bf v}_\alpha | \leq C || \curl H
  ||_{{\bf L}^2({\cal B}_\alpha)} || \curl {\bf v}_\alpha
||_{{\bf L}^2(\Omega)}.
\]
Since $H \in {\cal C}^{\infty}([0, T] \times \overline{\Omega})$
we have
\[
|| \curl H ||_{{\bf L}^2({\cal B}_\alpha)} \leq || \curl H
||_{L^\infty({\cal B}_\alpha)} \alpha^{d} (\sum_{j=1}^m
|B_j|)^{\frac{1}{2}} \leq C \alpha^{d},
\]
which gives
\[
\ds | \sum_{j=1}^m (\frac{1}{\eps_0}- \frac{1}{\eps_j}) \int_{z_j
+ \alpha B_j} \curl H\cdot \curl {\bf v}_\alpha | \leq C
\alpha^{d} || \curl {\bf v}_\alpha ||_{{\bf L}^2(\Omega)}
\]
and so, \be\label{rel-lem1-2} \ds \mu_0\partial_t \int_{\Omega}
\frac{1}{\eps_\alpha} |\curl {\bf v}_\alpha|^2 +
\partial_t \int_{\Omega} |H_\alpha - H|^2 \leq C \alpha^{d}
(\int_{\Omega} \frac{1}{\eps_\alpha} |\curl {\bf v}_\alpha|^2 +
\int_{\Omega} |H_\alpha - H|^2)^{1/2}. \ee From the Gronwall Lemma
it follows that \be \label{laast1} (\int_{\Omega}
\frac{1}{\eps_\alpha} |\curl {\bf v}_\alpha|^2)^{1/2} +
(\int_{\Omega} |H_\alpha - H|^2) )^{1/2} \leq C \alpha^{d}. \ee
Combining (\ref{laast1}) with the fact that
\[\ds
|| \partial_t (H_\alpha - H)||_{L^{\infty}(0, T; H^{-1}(\Omega))}
\leq C || \curl {\bf v}_\alpha ||_{L^{\infty}(0, T; {\bf
L}^2(\Omega))},
\]
the following estimate holds \be \label{estenergy1} \ds ||
H_\alpha - H||_{L^{\infty}(0, T; {\bf L}^2(\Omega))} + ||
\partial_t (H_\alpha - H)||_{L^{\infty}(0, T; {\bf L}^2(\Omega))}
\leq C \alpha^{d}. \ee Now, taking (formally) ${\bf v} =
\partial_t (H_\alpha - H)$ in (\ref{eq2}) we arrive at
\[
\ds \mu_0\partial_t \int_{\Omega} \Bigr[ |\partial_t (H_\alpha -
H)|^2 + \frac{1}{\eps_\alpha} |\curl (H_\alpha -H) |^2 \Bigr] = 2
\sum_{j=1}^m (\frac{1}{\eps_0} - \frac{1}{\eps_j}) \int_{z_j +
\alpha B_j} \curl H \cdot \curl
\partial_t (H_\alpha -H).
\]

By using the regularity of $H$ in $\Omega$ and estimate
(\ref{estenergy1}) given above, we see that
\[
\ds | \sum_{j=1}^m (\frac{1}{\eps_0} - \frac{1}{\eps_j}) \int_{z_j
+ \alpha B_j} \curl H \cdot \curl \partial_t (H_\alpha - H) | \leq
C ||\curl H ||_{{\bf H}^2({\cal B}_\alpha)} ||  \partial_t
(H_\alpha - H)||_{{\bf H}^{-1}(\Omega)} \leq C \alpha^{2d},
\]
where $C$ is independent of $t$ and $\alpha$, and so, we obtain
\[
\ds
\partial_t \int_{\Omega} \Bigr[
|\partial_t (H_\alpha - H)|^2 + \frac{1}{\eps_\alpha} |\curl
(H_\alpha -H) |^2 \Bigr] \leq C \alpha^{2d}
\]
which yields the following estimate
\[
|| \partial_t (H_\alpha - H) ||_{L^{\infty}(0, T; {\bf
L}^2(\Omega))} + || H_\alpha - H ||_{L^{\infty}(0, T;
X_N(\Omega))}  \leq C \alpha^{d},
\]
where $C$ is independent of $\alpha$ and the points $\{
z_j\}_{j=1}^m$.

\square\\

Now, we can estimate $\curl H_\alpha -\curl H$ as follows.
\begin{proposition}\label{est-prop1}
Let $H_\alpha$ and $H$ be solutions to the problems
(\ref{alphwam}) and (\ref{wom}) respectively. There exist
constants $0<\alpha_0$, $C$ such that for $0<\alpha<\alpha_0$ the
following estimate holds: \be\label{estenergy2} \ds || \curl
(H_\alpha -H)||_{L^{\infty}(0, T; {\bf L}^2(\Omega))}\leq C
\alpha^{d}, \ee
\end{proposition}
\proof To prove estimate (\ref{estenergy2}) it is useful to
introduce the following function

\be\label{r2} \ds \hat{v}(x) = \int_0^T v(x, t) z(t) \; dt \in
L^2(\Omega), \ee where $v \in L^1(0, T; L^2(\Omega))$ and $z(t)$
is a given function in
 ${\cal C}^\infty_0(]0, T[)$.\\
Then,
\[
\ds \hat{H}(x) = \int_0^T H(x, t) z(t) \; dt {\rm \; and \;}
  \hat{H}_\alpha(x) = \int_0^T H_\alpha(x, t) z(t) \; dt
\in X(\Omega),
\]
which by relation (\ref{estenergy0}) give
\[
\left\{
\begin{array}{l}
(\hat{H}_\alpha
- \hat{H})  \in {\bf H}^1(\Omega),\\
\nm \ds \curl  \curl  (\hat{H}_\alpha - \hat{H}) =  0(\alpha^{d})
\quad
{\rm in}\; \Omega,\\
\nm \dvg(\hat{H}_\alpha  - \hat{H})  = 0 \quad {\rm in}\;
\Omega,\\  \nm (\hat{H}_\alpha  - \hat{H}) \times \nu |_{\partial
\Omega} = 0,
\end{array}
\right.
\]
and so, \be\label{r3} ||\curl(\hat{H}_\alpha - \hat{H})||_{{\bf
L}^2(\Omega)}=O(\alpha^{d}). \ee The fact that $\curl(H_\alpha -
H)$ belongs to $L^\infty(0, T; {\bf L}^2(\Omega))$ and by using
(\ref{r2}) and (\ref{r3}) we arrive at:
\[
\int_{\Omega} |\curl H_\alpha(x,t) - \curl H(x,t)|^2~dx
=O(\alpha^{2d})\quad \mbox{ a.e. in }t\in (0,T),
\]
which means that
\[
||\curl (H_\alpha -H)||_{{\bf L}^2(\Omega)}=O(\alpha^{d})\quad
\mbox{ a.e. in }t\in (0,T).
\]
This equation can be bounded easily according to $t\in (0,T)$.
Thus, estimate (\ref{estenergy2}) holds. \square\\

Before formulating our main result in this section, let us denote
$\Phi_j, j=1, \ldots, m$ the unique vector-valued solution of the
following free space Laplace equation: \be \label{Phi} \left\{
\begin{array}{l} \Delta \Phi_j = 0 {\rm \; in\;} B_j, {\rm \; and \;}
\R^d \setminus \overline{B_j},\\ \nm \Phi_j {\rm \; is \;
continuous \; across\; } \partial B_j,\\ \nm \ds
\frac{\eps_j}{\eps_0} \frac{\partial \Phi_j}{\partial \nu_j}|_+ -
\frac{\partial \Phi_j}{\partial \nu_j}|_- = - \nu_j,\\ \nm \ds
\lim_{|y| \rightarrow + \infty} |\Phi_j(y)| = 0,
\end{array}
\right. \ee where $\nu_j$ denotes the outward unit normal to
$\partial B_j$, and superscripts $-$ and $+$ indicate the limiting
values as the point approaches $\partial B_j$ from outside $B_j$,
and from inside $B_j$, respectively. The existence and uniqueness
of this $\Phi_j$ can be established using single layer potentials
with suitably chosen densities, see \cite{CMV} for the case of
conductivity problem. For each inhomogeneity $z_j+ \alpha B_j$ we
introduce the  polarizability tensor $M_j$ which is a $d \times
d$, symmetric, positive definite matrix associated with the j-th
inhomogeneity, given by \begin{equation} \label{Meq} \ds (M_j)_{k,
l} = e_k \cdot (\int_{\partial B_j}
 (\nu_j +
(\frac{\eps_j}{\eps_0} - 1) \frac{\partial \Phi_j}{\partial
\nu_j}|_+ (y)) y \cdot e_l \; d\sigma_j(y)). \end{equation}
 Here $(e_1, \ldots, e_d)$ is an
orthonormal basis of $\R^d$. In terms of this function we are able
to prove the following result about the asymptotic behavior of
$\ds \curl H_\alpha \cdot\nu_j|_{\partial (z_j + \alpha B_j)^+}$.
\begin{theorem} \label{thm0} Suppose that (\ref{f1}) is satisfied and
 let $\Phi_j, j=1, \ldots, m$ be given as in (\ref{Phi}). Then, for the solutions
 $H_\alpha$, $H$ of problems (\ref{alphwam}) and (\ref{wom}) respectively, and for
  $y \in \partial
B_j$ we have \be \label{eq3} \ds(\curl H_\alpha (z_j + \alpha
y)\cdot\nu_j)|_{\partial (z_j + \alpha B_j)^+} = \curl H(z_j, t)
\cdot\nu_j \ee$$+ (1-\frac{\eps_j}{\eps_0})\frac{\partial
\Phi_j}{\partial \nu_j}|_+ (y) \cdot \curl H(z_j, t) +  o(1). $$
The term $o(1)$ uniform in $y \in \partial B_j$ and $t \in (0, T)$
and depends on the shape of $\{B_j \}_{j=1}^{m}$ and $\Omega$, the
constants $c_0$, $T$, $\eps_0$, $\{ \eps_j \}_{j=1}^{m}$, the data
$\varphi, \psi,$ and $f$, but is  otherwise independent of the
points $\{z_j \}_{j=1}^{m}$.
\end{theorem}
\proof

Let $\mathcal{H}_\alpha = \curl H_\alpha(x, t)$ and
$\mathcal{H}_0= \curl H(x, t)$. Then, according to
(\ref{wom})-(\ref{alphwam}) we have \be \label{u}
\mu_0\partial_t^2 H_\alpha - \curl \frac{1}{\eps_\alpha}
\mathcal{H}_\alpha = 0 {\rm \; and \;} \curl \mathcal{H}_\alpha =
0, {\rm \;  for \;}x\in\Omega. \ee We restrict, for simplicity,
our attention to the case of a single inhomogeneity, i.e., the
case $m=1$. The proof for any fixed number $m$ of well separated
inhomogeneities follows by iteration of the argument that we will
present for the case $m=1$. In order to further simplify notation,
we assume that the single inhomogeneity has the form $\alpha B$,
that is, we assume it is centered at the origin. We denote the
electromagnetic permeability inside $\alpha B$  by $\eps_*$ and
define $\Phi_*$ the same as $\Phi_j$, defined in (\ref{Phi}), but
with $B_j$ and $\eps_j$ replaced by $B$ and $\eps_*$,
respectively. Define $\nu$ to be the outward unit normal to
$\partial B$. Now, following a common practice in multiscale
expansions
 we introduce the local variable $\ds y = \frac{x}{\alpha}$, then the domain
 $\tilde{\Omega} = \ds(\frac{\Omega}{\alpha})$ is well defined.\\
Next, let $\varpi$ be given in
 ${\cal C}^\infty_0(]0, T[)$. For any function
  $v \in {\bf L}^1(0, T; {\bf L}^2(\Omega))$, we define
\[
\hat{v}(x) = \ds  \int_0^T v(x, t) \,  \varpi(t) \; dt \in {\bf
L}^2(\Omega).
\]
We remark that $\widehat{\partial_t v} (x) = - \ds \int_0^T v(x,
t) \varpi^\prime(t) \; dt.$ So that we deduce from (\ref{u}) that
$\hat{\mathcal{H}}_\alpha$ satisfies
\[
\left\{ \begin{array}{l} \ds
 \curl \frac{1}{\eps_\alpha} \hat{\mathcal{H}}_\alpha = \int_0^T H_\alpha \,
\varpi^{\prime \prime}(t) \; dt
\quad {\rm in\;} \Omega,\\
\nm \curl \hat{\mathcal{H}}_\alpha = 0 \quad {\rm in\;} \Omega.
\end{array}
\right.
\]
Analogously,  $\hat{\mathcal{H}}$ satisfies
\[
\left\{ \begin{array}{l} \ds
 \frac{1}{\eps_0}\curl \hat{\mathcal{H}} = \int_0^T H  \, \varpi^{\prime \prime}(t) \; dt
\quad {\rm in\;} \Omega,\\
\nm \curl \hat{\mathcal{H}}  = 0 \quad {\rm in\;} \Omega.
\end{array}
\right.
\]
Indeed, we have $\hat{\mathcal{H}}_\alpha \times \nu =
\hat{\mathcal{H}} \times \nu = \curl_{\partial \Omega} \hat{f}
\times \nu$ on the boundary $\partial \Omega$, where
$\curl_{\partial \Omega}$ is the tangential curl. Following
\cite{AVV} and \cite{A}, we introduce $q_\alpha^*$ as the unique
solution to the following problem
\[
\left\{ \begin{array}{l} \ds
 \Delta q_\alpha^* = 0
\quad {\rm in\;} \tilde{\Omega} = (\frac{\Omega}{\alpha})
\setminus \overline{B}
{\rm \; and \; in\;} B, \nonumber\\
\nm \ds
q_\alpha^* {\rm \;is\;continuous\;across\;} \partial B, \nonumber\\
\nm \ds \eps_0 \frac{\partial q_\alpha^*}{\partial \nu}|_+ -
\eps_*
 \frac{\partial q_\alpha^*}{\partial \nu}|_{-} = - (\eps_0 - \eps_*)
\hat{\mathcal{H}}(\alpha y) \cdot \nu \quad {\rm on\;} \partial B,\\
\nm \ds q_\alpha^* = 0 \quad {\rm on\;} \partial \tilde{\Omega}.
\nonumber
\end{array}
\right.
\]
The jump condition
$$
\ds \eps_0 \frac{\partial q_\alpha^*}{\partial \nu}|_+ - \eps_*
 \frac{\partial q_\alpha^*}{\partial \nu}|_{-} = - (\eps_0 - \eps_*)
\hat{\mathcal{H}}(\alpha y) \cdot \nu \quad {\rm on\;} \partial B
$$
guarantees that $\hat{\mathcal{H}}_\alpha(x) -
\hat{\mathcal{H}}(x) - \grad_y q_\alpha^* (\frac{x}{\alpha})$
belongs to the functional space $X_N(\Omega)$, where
$\grad_{\partial \Omega}$ is the tangential gradient. Since
\[
\left\{ \begin{array}{l} \ds
 \curl \frac{1}{\eps_\alpha}( \hat{\mathcal{H}}_\alpha
- \hat{\mathcal{H}} - \grad_y q_\alpha^* (\frac{x}{\alpha}) ) =
\int_0^T \Bigr[ H_\alpha - \chi(\Omega \setminus \overline{\alpha
B}) H + \frac{\eps_*}{\eps_0} \chi(\alpha B) H \Bigr]
\varpi^{\prime \prime}(t) \; dt
\quad {\rm in\;} \Omega,\\
\nm \ds \curl ( \hat{\mathcal{H}}_\alpha - \hat{\mathcal{H}} -
\grad_y q_\alpha^* (\frac{x}{\alpha}))
= 0 \quad {\rm in\;} \Omega,\\
\nm ( \hat{\mathcal{H}}_\alpha - \hat{\mathcal{H}} - \grad_y
q_\alpha^* (\frac{x}{\alpha}) ) \times \nu = 0 \quad {\rm on\;}
\partial \Omega,
\end{array}
\right.
\]
where $\chi(\omega)$ is the characteristic function of the domain
$\omega$, we arrive, as a consequence of the energy estimate given
by Lemma \ref{est-lem1}, at the following
\[
\left\{ \begin{array}{l}
 (\hat{\mathcal{H}}_\alpha
- \hat{\mathcal{H}} - \grad_y q_\alpha^* (\frac{x}{\alpha}))
\in X_N(\Omega),\\
\nm \ds
 \curl \frac{1}{\eps_\alpha} ( \hat{\mathcal{H}}_\alpha
- \hat{\mathcal{H}} - \curl_y q_\alpha^* (\frac{x}{\alpha})) =
0(\alpha)
\quad {\rm in\;} \Omega,\\
\nm \ds \curl ( \hat{\mathcal{H}}_\alpha - \hat{\mathcal{H}} -
\grad_y q_\alpha^* (\frac{x}{\alpha}))
= 0 \quad {\rm in\;} \Omega,\\
\nm ( \hat{\mathcal{H}}_\alpha - \hat{\mathcal{H}} - \grad_y
q_\alpha^* (\frac{x}{\alpha}))  \times \nu = 0 \quad {\rm on\;}
\partial \Omega.
\end{array}
\right.
\]
From \cite{AVV} we know that this yields the following estimate
\[
|| \curl \frac{1}{\eps_\alpha}( \hat{\mathcal{H}}_\alpha -
\hat{\mathcal{H}} - \grad_y q_\alpha^* (\frac{x}{\alpha}) )
||_{L^2(\Omega)} + ||  \hat{\mathcal{H}}_\alpha -
\hat{\mathcal{H}} - \grad_y q_\alpha^* (\frac{x}{\alpha})
||_{L^2(\Omega)} \leq C \alpha,
\]
and so,
\[
( \hat{\mathcal{H}}_\alpha - \hat{\mathcal{H}} - \grad_y
q_\alpha^* (\frac{x}{\alpha})) \cdot \nu |_+ = 0(\alpha) \quad
{\rm on\;}
\partial (\alpha B).
\]
Now, we denote by $q_*$ be the unique (scalar) solution to
\[
\left\{ \begin{array}{l} \ds
 \Delta q_* = 0
\quad {\rm in\;} \RR^d \setminus \overline{B}
{\rm \; and \; in\;} B,\\
\nm \ds
q_* {\rm \;is\;continuous\;across\;} \partial B, \\
\nm \ds \eps_0 \frac{\partial q_*}{\partial \nu}|_+ - \eps_*
 \frac{\partial q_*}{\partial \nu}|_{-} = - (\eps_0 - \eps_*)
\hat{\mathcal{H}}(0) \cdot \nu \quad {\rm on\;} \partial B,\\
\nm \ds \lim_{|y| \rightarrow + \infty} q_* = 0.
\end{array}
\right.
\]
In the spirit of Theorem 1 in \cite{CMV} it follows that
\[
\ds ||(\grad_y q_* - \grad_y q_\alpha^*)(\frac{x}{\alpha})
||_{{\bf L}^2(\Omega)} \leq C \alpha^{1/2},
\]
which yields
\[
( \hat{\mathcal{H}}_\alpha - \hat{\mathcal{H}} - \grad_y q_*
(\frac{x}{\alpha})) \cdot \nu = o(1) \quad {\rm on\;} \partial
(\alpha B).
\]
Writing $q_*$ in terms of $\Phi_*$ gives
\[
\ds \int_0^T \Bigr[ (\curl H_\alpha(\alpha y)\cdot\nu)|_{\partial
(\alpha B)^+}  -  \nu \cdot \curl H(0, t) - (\frac{\eps_0}{\eps_*}
- 1) \frac{\partial \Phi_*}{\partial \nu}|_+ (y) \cdot \curl H(0,
t) \Bigr] \varpi(t) \; dt =  o(1),
\]
for any $\varpi \in {\cal C}^\infty_0(]0, T[)$, and so, by
iterating the same argument for the case of $m$ (well separated)
inhomogeneities $z_j + \alpha B_j, j=1, \ldots, m$, we arrive at
the promised asymptotic formula (\ref{eq3}).

\square

%%%%%%%%%%%%%%%%%%%%%%%%%%%%%%%%%%%%%%%%%%%%%%%%%%%%%%%%%%%%%%%%%%%%%%%%%%%%%%%%%%%%%%%%
\section{Reconstruction} Before describing our
identification and reconstruction procedure, let us introduce the
following cutoff function $\beta(x)  \in {\cal
C}^{\infty}_0(\Omega)$ such that $\beta \equiv 1$ in a subdomain
$\Omega^{\prime}$ of $\Omega$ that contains the inhomogeneities
${\cal B}_\alpha$ and let $\eta \in \RR^d$. We will take in what
follows $H(x, t) = \eta^{\perp} e^{i \eta \cdot x
-i\sqrt{\eps_0}|\eta | t}$ where $\eta^{\perp}$ is a unit vector
that is orthogonal to $\eta$ which corresponds to taking
$\varphi(x) = \eta^{\perp} e^{i \eta \cdot x}, \psi(x) = - i
\sqrt{\eps_0}|\eta| \eta^{\perp} e^{i \eta \cdot x}, $ and $f(x,
t) = \eta^{\perp}\times \nu e^{i \eta\cdot x - i
\sqrt{\eps_0}|\eta| t}$ and assume that we are in possession of
the measurements of:
$$\curl H_\alpha \times \nu \quad \mbox{ on } \Gamma \times (0, T),$$
 where $\Gamma$ is an open part of $\partial \Omega$. Suppose now that $T$ and
the part $\Gamma$ of the boundary $\partial \Omega$ are such that
they  geometrically control $\Omega$ which means that they satisfy
the geometric control hypothesis of the work of Bardos, Lebeau and
Rauch in \cite{BLR}:

\begin{definition}
Let $\Gamma$ be an open subset of $\partial\Omega$ and $T$ a
positive number. One says that $(\Gamma,T)$ geometrically control
$\Omega$ if for every geometrical optic ray $s\mapsto \gamma(s)$,
there exists $s_0\in ]0,T[$ such that $\gamma(s_0)\in \Gamma\times
]0,T[$ and $\gamma(s_0)$ non diffractive point.
\end{definition}

It follows from \cite{NI} (see also \cite{LAG}, \cite{K} and
\cite{KO}) that we can construct (a unique) $g_\eta \in H^1_0(0,
T; TL^2(\Gamma))$ (by the Hilbert Uniqueness Method) such that the
unique weak solution $w_\eta$ to \be \label{wetam} \left\{
\begin{array}{l}
\ds (\partial_t^2 + \curl  \curl ) w_\eta  = 0 \quad
{\rm in}\; \Omega \times (0, T),\\
\nm \dvg w_\eta = 0 \quad
{\rm in}\; \Omega \times (0, T),\\
\nm w_\eta  |_{t=0} = \beta(x) \eta^{\perp} e^{i \eta \cdot x},
\partial_t w_\eta  |_{t=0} = 0 \quad {\rm in}\; \Omega,\\ \nm
w_\eta  \times \nu |_{\partial \Omega \setminus \overline{\Gamma}
 \times (0, T)} = 0,\\
\nm w_\eta  \times \nu |_{\Gamma
 \times (0, T)} = g_\eta,
\end{array}
\right. \ee satisfies $w_\eta(T) = \partial_t w_\eta (T) = 0$ in
$\Omega$.

Let $\theta_\eta \in H^1(0, T; TL^2(\Gamma))$ denote the unique
solution of the Volterra equation of second kind \be\label{eq4m}
\left\{\begin{array}{l} \ds
\partial_t \theta_\eta (x, t) + \int_t^T e^{- i | \eta | (s -t)}
( \theta_\eta (x, s) - i | \eta | \partial_t \theta_\eta (x, s))
\; ds = g_\eta(x, t) \quad {\rm for\;} x \in \Gamma, t \in (0, T),
\\ \theta_\eta(x, 0) = 0 \quad {\rm for\;} x \in \Gamma.
\end{array}
\right. \ee The existence and uniqueness of this $\theta_\eta$ in
${\bf H}^1(0, T; TL^2(\Gamma))$ for any $\eta \in \RR^d$ can be
established using the resolvent kernel. However, observing from
differentiation of (\ref{eq4m}) with respect to $t$ that
$\theta_\eta$ is the unique solution of the ODE: \be \label{eq4p}
\left\{
\begin{array}{l}
\ds
\partial_t^2 \theta_\eta - \theta_\eta = e^{i |\eta| t}
\partial_t ( e^{-i |\eta| t} g_\eta) \quad {\rm for\;} x \in
\Gamma, t \in (0, T), \\ \theta_\eta(x, 0) = 0,
\partial_t \theta_\eta(x, T) = 0  \quad {\rm for\;} x \in \Gamma,
\end{array}
\right. \ee the function $\theta_\eta$ may be find (in practice)
explicitly with variation of parameters and it also immediately
follows from this observation that $\theta_\eta$ belongs to
${\bf H}^2(0, T; TL^2(\Gamma))$.\\
We introduce $v_{ \eta}$ as the unique weak solution (obtained by
transposition as done in \cite{LM} and in \cite{L} [Theorem 4.2,
page 46] for the scalar function) in $ {\cal C}^0(0, T; X(\Omega))
\cap {\cal C}^1(0, T; L^2(\Omega))$ to the following problem
\[
\left\{
\begin{array}{l}
\ds (\partial_t^2 + \curl  \curl ) v_\eta  = 0 \quad
{\rm in}\; \Omega \times (0, T),\\
\nm \dvg v_\eta = 0 \quad
{\rm in}\; \Omega \times (0, T),\\
\nm v_\eta  |_{t=0} = 0  \quad {\rm in}\; \Omega,\\
\nm \ds
\partial_t v_\eta  |_{t=0} = \sum_{j=1}^m i (1 - \frac{\eps_0}{\eps_j}) \eta \times
 (\nu_j +
(\frac{\eps_0}{\eps_j} - 1) \frac{\partial \Phi_j}{\partial
\nu_j}|_+) e^{i \eta \cdot z_j} \delta_{\partial (z_j + \alpha
B_j)}  \in Y(\Omega) \quad {\rm in}\; \Omega,\\ \nm v_\eta  \times
\nu |_{\partial \Omega
 \times (0, T)} = 0.
\end{array}
\right.
\]
%%%%%%%%%%%%%%%%%%%%%%%%%%%%%%%%%%
Then, the following holds.

\begin{proposition}\label{p4.1}
Suppose that $\Gamma$ and $T$ geometrically control $\Omega$. For
any $\eta \in \RR^d$ and $\eta^{\perp}$ unit vector in $\RR^d$
that is orthogonal to $\eta$, we have \be\label{rel-prop3.1}
\int_0^T \int_\Gamma g_\eta \cdot(\curl
v_{\eta}\times\nu)=\alpha^{d} \sum_{j=1}^m  \eps_0(1 -
\frac{\eps_j}{\eps_0}) e^{2 i \eta \cdot z_j} \Big(\eta \times
(\int_{\partial B_j}
 (\nu_j \ee \[+
(\frac{\eps_j}{\eps_0} - 1) \frac{\partial \Phi_j}{\partial
\nu_j}|_+ (y)) ))y\cdot\eta\Big)\cdot \eta^{\perp} \; ds_j(y)+
O(\alpha^{d}).
\]
\end{proposition}
\proof Multiply the equation $\ds (\partial_t^2 + \curl  \curl )
v_\eta  = 0$ by $w_\eta$ and integrating by parts in $t\in(0,T)$,
we get
\[
\ds \int_0^T \int_\Omega (\partial_t^2 +\curl \curl ) v_{\eta}
w_\eta =\int_0^T \int_\Omega \curl \curl v_{\eta}
w_\eta+\int_\Omega\int_0^T\partial_t^2 v_{\eta} w_\eta
\]
\[=\int_0^T \int_\Omega \curl \curl v_{\eta}
w_\eta+\int_\Omega\partial_t v_{\eta} w_\eta|_{t=0}-\partial_t
v_{\eta} w_\eta|_{t=T}-\int_\Omega\int_0^T\partial_t v_{\eta}
\partial_t w_\eta \]

\[=\int_0^T \int_\Omega \curl \curl v_{\eta}
w_\eta+\int_\Omega\partial_t v_{\eta} w_\eta|_{t=0}+\int_\Omega
v_{\eta} \partial_t w_\eta|_{t=0}+\int_\Omega\int_0^T v_{\eta}
\partial_t^2 w_\eta.
\]
So, by Green's formula,
\[
\ds \int_0^T \int_\Omega (\partial_t^2 +\curl \curl ) v_{\eta}
w_\eta = -\alpha^{d-1} \sum_{j=1}^m i (1 - \frac{\eps_j}{\eps_0})
e^{2 i \eta \cdot z_j} \eta \times( \int_{\partial B_j}
 (\nu_j +\]
 \[
(\frac{\eps_j}{\eps_0} - 1) \frac{\partial \Phi_j}{\partial
\nu_j}|_+ (y)) e^{i \alpha \eta \cdot y})\cdot \beta(y)
\eta^{\perp} \; ds_j(y)
\]
\[-
\eps_{0}^{-1} \int_0^T \int_\Gamma g_\eta \cdot(\curl v_{
\eta}\times\nu) = 0.
\]
Therefore
\[
\alpha^{d-1} \sum_{j=1}^m i (1 - \frac{\eps_j}{\eps_0}) e^{2 i
\eta \cdot z_j} \eta \times (\int_{\partial B_j}
 (\nu_j +
(\frac{\eps_j}{\eps_0} - 1) \frac{\partial \Phi_j}{\partial
\nu_j}|_+ (y)) e^{i \alpha \eta \cdot y} )\cdot \beta(y)
\eta^{\perp}\; ds_j(y)  = \] \[- \eps_{0}^{-1} \int_0^T
\int_\Gamma g_\eta \cdot(\curl v_{ \eta}\times\nu).
\]
Now, we take the Taylor expansion of $\alpha^{d-1} e^{i \alpha
\eta \cdot y}$ in the left side of the last equation and we use
the definition of the cutoff function $\beta(x)$, we obtain the
convenient asymptotic formula (\ref{rel-prop3.1}).

\square

 To identify the locations and certain properties of
the small inhomogeneities ${\cal B}_\alpha$ let us view the
averaging of the boundary measurements \[\ds \curl H_\alpha \times
\nu |_{\Gamma \times (0, T)},\] using the solution $\theta_\eta$
to the Volterra equation (\ref{eq4m}) or equivalently the ODE
(\ref{eq4p}), as a function of $\eta$. The
following holds.\\

\begin{theorem} \label{th2}
Let $\eta \in \RR^d$ and $\eta^{\perp}$ be a unit vector in
$\RR^d$ that is orthogonal to $\eta$. Let $H_\alpha$ be the unique
solution in $ {\cal C}^0(0, T; X(\Omega)) \cap {\cal C}^1(0, T;
L^2(\Omega))$ to the Maxwell's equations (\ref{alphwam}) with $
\varphi(x) = \eta^{\perp} e^{i \eta \cdot x},$ $ \psi(x) = - i
\sqrt{\eps_0}| \eta | \eta^{\perp} e^{i \eta \cdot x},$ and $ f(x,
t) = \eta^{\perp} e^{i \eta \cdot x  -  i \sqrt{\eps_0}| \eta |
t}.$ Suppose that $\Gamma$ and $T$ geometrically control $\Omega$,
then we have \be \label{eqam}
\begin{array}{l}
\ds \int_0^T \int_\Gamma \Bigr[ \theta_\eta \cdot (\curl H_\alpha
\times \nu - \curl H \times \nu) +
\partial_t \theta_\eta \cdot \partial_t (\curl H_\alpha \times \nu
- \curl H \times \nu)  \Bigr] =  \\ \nm \ds \alpha^d \sum_{j=1}^m
(\eps_0 - \eps_j) e^{2 i \eta \cdot z_j} (\eta\times M_j(\eta))
\cdot \eta^{\perp} \;
  + O(\alpha^d),
  \end{array}
\ee where $\theta_\eta$ is the unique solution to the Volterra
equation (\ref{eq4p}) with $g_\eta$ defined as the boundary
control  in (\ref{wetam}) and $M_j$ is the polarization tensor of
$B_j$, defined by \be \label{pt} \ds (M_j)_{k, l} = e_k \cdot
(\int_{\partial B_j}
 (\nu_j +
(\frac{\eps_j}{\eps_0} - 1) \frac{\partial \Phi_j}{\partial
\nu_j}|_+ (y)) y \cdot e_l \; ds_j(y)). \ee Here $(e_1,\cdots,
e_d)$ is an orthonormal basis of $\RR^d$. The term $O(\alpha^d)$
is independent of the points $\{z_j,\quad j=1,\cdots, m\}$.
\end{theorem}
\proof From $\partial_t \theta_\eta (T) = 0$ and $(\curl H_\alpha
\times \nu - \curl H \times \nu)|_{t=0}  = 0$ the term $ \ds
\int_0^T \int_\Gamma
\partial_t \theta_\eta \cdot\partial_t (\curl H_\alpha
\times \nu - \curl H \times \nu)$ has to be interpreted as follows
\be \label{d} \ds \int_0^T \int_\Gamma
\partial_t \theta_\eta\cdot \partial_t (\curl H_\alpha
\times \nu - \curl H \times \nu)  = - \int_0^T \int_\Gamma
\partial^2_t \theta_\eta \cdot(\curl H_\alpha
\times \nu - \curl H \times \nu). \ee Next, introduce
\be\label{def1} \ds \tilde{H}_{\alpha, \eta}(x, t) = H(x, t) +
\int_0^t e^{- i \sqrt{\eps_0} | \eta| s} v_{\eta}(x, t-s)\; ds, x
\in \Omega, t \in (0, T). \ee We have
\[\begin{array}{l}
\ds \int_0^T \int_\Gamma \Bigr[ \theta_\eta \cdot(\curl H_\alpha
\times \nu - \curl H \times \nu ) +
\partial_t \theta_\eta\cdot \partial_t (\curl H_\alpha
\times \nu - \curl H \times \nu ) \Bigr]
= \\
\nm \ds \int_0^T \int_\Gamma \Bigr[ \theta_\eta \cdot(\curl
H_\alpha \times \nu- \curl \tilde{H}_{\alpha,\eta}\times \nu) +
\partial_t \theta_\eta\cdot \partial_t (\curl
H_\alpha \times \nu- \curl \tilde{H}_{\alpha,\eta}\times \nu ) \Bigr]\\
\nm \ds + \int_0^T \int_{\Gamma} \Bigr[ \theta_\eta\cdot \int_0^t
e^{- i \sqrt{\eps_0} | \eta| s} v_{\eta}(x, t-s)\times\nu\; ds +
\partial_t \theta_\eta \cdot\partial_t \int_0^t e^{- i
\sqrt{\eps_0} | \eta| s} v_{\eta}(x, t-s)\times\nu\; ds \Bigr].
\end{array}
\]
Since $\theta_\eta$ satisfies the Volterra equation (\ref{eq4p})
and
\[\begin{array}{l}
\ds \partial_t ( \int_0^t e^{-  i \sqrt{\eps_0} | \eta | s}
v_{\eta}(x, t-s)\times\nu\; ds ) =
\partial_t (e^{- i \sqrt{\eps_0} |\eta| t}  \int_0^t e^{ i \sqrt{\eps_0}
| \eta | s} v_{\eta}(x, s)\times\nu\; ds )
\\ \nm \ds =  i \sqrt{\eps_0} |\eta| e^{- i \sqrt{\eps_0} |\eta| t}
 \int_0^t e^{ i \sqrt{\eps_0}| \eta | s}
v_{\eta}(x, s)\times\nu\; ds + v_{\eta}(x, t)\times\nu,
\end{array}
\]
we obtain by integrating by parts over $(0, T)$ that
\[\begin{array}{l}
  \ds \int_0^T \int_{\Gamma} \Bigr[
\theta_\eta\cdot \int_0^t e^{-  i\sqrt{\eps_0}  | \eta | s}
v_{\eta}(x, t-s)\times\nu\; ds +
\partial_t \theta_\eta\cdot
\partial_t \int_0^t e^{- i\sqrt{\eps_0} | \eta | s} v_{\eta}(x, t-s)\times\nu\;
 ds \Bigr]  \\= \ds
\int_0^T \int_{\Gamma} ( v_{\eta}(x, t)\times\nu)\cdot (\partial_t
\theta_\eta + \int_t^T \theta_\eta(s) e^{i \sqrt{\eps_0} |\eta|
(t-s)} \;ds)\\ \nm \quad \ds  - i \sqrt{\eps_0} |\eta| (e^{- i
\sqrt{\eps_0}|\eta| t}
\partial_t \theta_\eta (t))\cdot \int_0^t e^{ i\sqrt{\eps_0} | \eta | s}
v_{\eta}(x, s)\times\nu\; ds\; dt\\ \nm = \ds \int_0^T
\int_{\Gamma} v_{\eta}(x, t)\times\nu\cdot (\partial_t \theta_\eta
+ \int_t^T (\theta_\eta(s) - i \sqrt{\eps_0} |\eta|
\partial_t \theta_\eta(s))
 e^{ i \sqrt{\eps_0}| \eta |(t- s)} \;ds) \;dt\\
\nm \ds
 = \int_0^T \int_{\Gamma} g_\eta(x, t)
\cdot (\curl v_{\eta}(x, t)\times\nu)\; dt
\end{array}
\]
and so, from Proposition \ref{p4.1} we obtain
\[\begin{array}{l}
\ds \int_0^T \int_\Gamma
 \Bigr[
\theta_\eta \cdot(\curl H_\alpha \times \nu - \curl H \times \nu )
+
\partial_t \theta_\eta \cdot\partial_t (\curl H_\alpha
\times \nu - \curl H \times \nu ) \Bigr]
= \\
\nm \ds \alpha^d  \sum_{j=1}^m  (1 - \frac{\eps_j}{\eps_0}) e^{2 i
\eta \cdot z_j} \big(\eta \times (\int_{\partial B_j}
 (\nu_j +
(\frac{\eps_j}{\eps_0} - 1) \frac{\partial \Phi_j}{\partial
\nu_j}|_+ (y)) \eta \cdot y)\big)\cdot \eta^{\perp} \; ds_j(y)\\
\nm \ds + \int_0^T \int_\Gamma
 \Bigr[
\theta_\eta \cdot(\curl H_\alpha \times \nu - \curl
\tilde{H}_{\alpha, \eta} \times \nu ) +
\partial_t \theta_\eta \cdot\partial_t (\curl H_\alpha \times \nu\\
\nm \ds  - \curl \tilde{H}_{\alpha, \eta} \times \nu ) \Bigr] +
o(\alpha^d).
\end{array}
\]
In order to prove Theorem {\ref{th2} it suffices then to show that
\be \label{p}
 \int_0^T \int_\Gamma
 \Bigr[
\theta_\eta \cdot(\curl H_\alpha \times \nu - \curl
\tilde{H}_{\alpha, \eta} \times \nu ) +
\partial_t \theta_\eta\cdot \partial_t (\curl H_\alpha \times \nu - \curl
\tilde{H}_{\alpha, \eta} \times \nu ) \Bigr] = o(\alpha^d). \ee
Since
\[
\left\{
\begin{array}{l}
\ds (\partial_t^2 - \curl\frac{1}{\eps_0}\curl) ( \int_0^t e^{- i
\sqrt{\eps_0} | \eta| s} v_{ \eta}(x, t-s)\; ds)
\\ \ds = \sum_{j=1}^m i (1 - \frac{\eps_j}{\eps_0}) \eta \times
(\nu_j + (\frac{\eps_j}{\eps_0} - 1) \frac{\partial
\Phi_j}{\partial \nu_j}|_+ (y)) e^{i \eta \cdot z_j}
\delta_{\partial (z_j + \alpha B_j)} e^{- i \sqrt{\eps_0} |\eta|
t} \quad {\rm in}\; \Omega \times (0, T),\\ \nm \ds ( \int_0^t
e^{- i \sqrt{\eps_0} | \eta| s} v_{
 \eta}(x, t-s)\; ds) |_{t=0} = 0,
\partial_t ( \int_0^t e^{- i \sqrt{\eps_0} | \eta| s} v_{ \eta}(x, t-s)\; ds)|_{t=0} = 0
 \quad {\rm in}\; \Omega,\\ \nm
\ds ( \int_0^t e^{- i \sqrt{\eps_0} | \eta| s} v_{\eta}(x, t-s)\;
ds)\times\nu |_{\partial \Omega \times (0, T)} = 0,
\end{array}
\right.
\]
it follows from Theorem \ref{thm0} that
\[
\left\{
\begin{array}{l}
\ds (\partial_t^2 - \curl\frac{1}{\eps_0}\curl) (H_\alpha -
\tilde{H}_{\alpha, \eta})  = O(\alpha^d)\in Y(\Omega), \quad {\rm
in}\; \Omega \times (0, T),\\ \nm (H_\alpha - \tilde{H}_{\alpha,
\eta}) |_{t=0} = 0,
\partial_t (H_\alpha - \tilde{H}_{\alpha, \eta}) |_{t=0} = 0
 \quad {\rm in}\; \Omega,\\ \nm
(H_\alpha - \tilde{H}_{\alpha, \eta})\times\nu|_{\partial \Omega
\times (0, T)} = 0.
\end{array}
\right.
\]
Following the proof of Proposition \ref{est-prop1}, we immediately
obtain
\[
|| H_\alpha - \tilde{H}_{\alpha, \eta} ||_{{\bf L}^2(\Omega)} =
O(\alpha^d), t \in (0, T), x \in \Omega,
\]
where $O(\alpha^d)$, $d=2,3$ is independent of the points $\{
z_j\}_{j=1}^m$. To prove (\ref{p}) it suffices then from (\ref{d})
to show that the  following estimate holds
\[
\ds || \curl H_\alpha \times \nu - \curl \tilde{H}_{\alpha, \eta}
\times \nu||_{L^2(0, T; TL^2(\Gamma))} = O(\alpha^d).
\]
Let
 $\theta$ be given in
 ${\cal C}^\infty_0(]0, T[)$ and define
\[
\ds\hat{\tilde{H}}_{\alpha, \eta}(x) = \ds  \int_0^T
\tilde{H}_{\alpha, \eta}(x, t) \theta(t) \; dt
\]
and
\[
\ds\hat{H}_{\alpha}(x) = \ds  \int_0^T H_{\alpha}(x, t) \theta(t)
\; dt.
\]
From definition (\ref{def1}) we can write \be \label{r8} \left\{
\begin{array}{l}
(\hat{H}_\alpha
- \hat{\tilde{H}}_\alpha)  \in {\bf H}^1(\Omega),\\
\nm \ds \curl  \curl  (\hat{H}_\alpha - \hat{\tilde{H}}_\alpha) =
O(\alpha^d) \in Y(\Omega) \quad
{\rm in}\; \Omega,\\
\nm \dvg(\hat{H}_\alpha  - \hat{\tilde{H}}_\alpha)  = 0 \quad {\rm
in}\; \Omega,\\  \nm (\hat{H}_\alpha  - \hat{\tilde{H}}_\alpha)
\times \nu |_{\partial \Omega} = 0.
\end{array}
\right. \ee In the spirit of the standard elliptic regularity
\cite{E} we deduce for the boundary value problem (\ref{r8}) that
\[\ds || \curl(\hat{H}_\alpha - \hat{\tilde{H}}_{\alpha})\times \nu
||_{{\bf L}^2(\Gamma)} = O(\alpha^d), \] for all $\theta \in {\cal
C}^\infty_0(]0, T[)$; whence
$$
||\curl(H_\alpha - \hat{H}_{\alpha})\times \nu||_{{\bf
L}^2(\Gamma)}
 = O(\alpha^d) {\rm \; a.\, e.\; in \;} t \in (0, T), $$
and so, the desired estimate (\ref{eqam}) holds. The proof of
Theorem \ref{th2} is
then over. \square\\
%%%%%%%%%%%%%%%%%%%%%%%%%%%%%%%%%%%%%%%%%%%%%%%%%%%%%%%%%%%%%%%%%%%%%%%%%%%%%%%%%%

Our identification procedure is deeply based on Theorem \ref{th2}.
 Let us neglect the asymptotically small remainder in the asymptotic formula
(\ref{eqam}), and define $\aleph_\alpha(\eta)$ by
\[
\ds \aleph_\alpha(\eta) =  \int_0^T \int_\Gamma
 \Bigr[
\theta_\eta \cdot(\curl(H_\alpha - H)\times \nu) +
\partial_t \theta_\eta \cdot\partial_t (\curl(H_\alpha - H)\times \nu ) \Bigr].
\]
Recall that the function $e^{2 i \eta \cdot z_j}$ is exactly the
Fourier Transform (up to a multiplicative constant) of the Dirac
function $\delta_{-2 z_j}$ (a point mass located at $- 2 z_j$).
From Theorem \ref{th2} it follows that the function $e^{2 i \eta
\cdot z_j}$ is (approximately) the Fourier Transform of a linear
combination of derivatives of point masses, or
\[
 \breve{\aleph}_\alpha(\eta) \approx \alpha^d \sum_{j=1}^m L_j \delta_{-2 z_j},
\]
where $L_j$ is a second order constant coefficient, differential
operator whose
 coefficients depend on the polarization tensor $M_j$ defined by (\ref{pt})
(see  \cite{CMV} for its properties) and $
\breve{\aleph}_\alpha(\eta)$ represents the inverse Fourier
Transform of $\aleph_\alpha(\eta)$.

The method of reconstruction consists in sampling values of
  $ \breve{\aleph}_\alpha(\eta)$ at some discrete set of points and then calculating
  the corresponding discrete inverse Fourier Transform. After a rescaling
  the support of this discrete inverse Fourier Transform yields the
location of the small inhomogeneities ${\cal B}_\alpha$. Once the
locations are known we may calculate the polarization tensors
$(M_j)^m_{j=1}$ by solving an appropriate linear system arising
from (\ref{eqam}).  This procedure generalizes the approach
developed in \cite{AMV} for the two-dimensional (time-independent)
inverse conductivity problem and generalize the results in
\cite{A} to the full time-dependent Maxwell's equations.

%%%%%%%%%%%%%%%%%%%%%%%%%%%%%%%%%%%%%%%%%%%%%%%%%%%%%%%%%%%%%%%%%%%%%%%%%%%%%%%%%%%%%%%

\end{document}